\documentclass[superscriptaddress,preprint]{revtex4}
\usepackage{mathrsfs}
\usepackage{amssymb}
\usepackage[tbtags]{amsmath}
\usepackage{graphicx}
\usepackage{epsfig,graphicx,times}
\usepackage{graphicx}
\usepackage{subfigure}
\usepackage{color}

\begin{document}
\title{Quantum computation with surface-state electrons by rapid population passages}
\author{X. Shi}
\affiliation{Quantum Optoelectronics Laboratory, School of Physics
and Technology, Southwest Jiaotong University, Chengdu 610031,
China}
\affiliation{Centre for Quantum Technologies and Department
of Physics, National University of Singapore, 3 Science Drive 2,
Singapore 117542}
\author{L. F. Wei\footnote{weilianfu@gmail.com}}
\affiliation{Quantum
Optoelectronics Laboratory, School of Physics and Technology,
Southwest Jiaotong University, Chengdu 610031, China}
\affiliation{State Key Laboratory of Optoelectronic Materials and
Technologies, School of Physics Science and Engineering, Sun Yet-sen
University, Guangzhou 510275, China}
\author{C. H. Oh\footnote{phyohch@nus.edu.sg}}
\affiliation{Centre for Quantum Technologies and Department of
Physics, National University of Singapore, 3 Science Drive 2,
Singapore 117542}
\date{\today}
\begin{abstract}
Quantum computation requires coherently controlling the evolutions of qubits. Usually, these
manipulations are implemented by precisely designing the durations
(such as the $\pi$-pulses) of the Rabi oscillations and tunable
interbit coupling. Relaxing this requirement, here we show that the
desired population transfers between the logic states can be
deterministically realized (and thus quantum computation could be
implemented) both adiabatically and non-adiabatically, by performing
the duration-insensitive quantum manipulations. Our proposal is
specifically demonstrated with the surface-state of electrons
floating on the liquid helium, but could also be applied to the
other artificially controllable systems for quantum computing.


\end{abstract}
\maketitle
\section{Introduction}
With the advent of the integrated circuits technology, the
components of the classical computer become smaller and smaller and inevitably quantum effects such as quantum interference have
to be met. Due to these quantum effects, the usual (classical)
computing technology might be invalid and computation based on the quantum laws opens a new chapter to overcome the limits of classical computing~\cite{Lloyds}. Indeed, it has been
shown that quantum computation could solve effectively certain
problems (e.g., factoring large numbers~\cite{Shor} and searching
databases~\cite{Grover}), which could not be solved by classical
algorithms. Different from the
bits in classical computer, the qubits in a quantum computer are
made by the controlled two-level quantum systems and could be
prepared at the superpositions of their basis states~\cite{Peter}. Therefore, quantum computer
provides an automatically-parallel computing and thus is
much more powerful than the
classical computer.

Basically, the central task for quantum computing is to coherently
control the transitions between the quantum logic states for
implementing various gate operations.
Generally, there are two well-known approaches to realize the
population transfers between two quantum states; one makes use
of the Rabi oscillations and the other is based on population
passages~\cite{Fleischhauer,Shore}. 
In the first way a constant-intensity resonant driving is applied to
the two-state system, and the population of the target state can
undergo a periodic variation with the Rabi frequency. This
implies that, by properly choosing the pulse duration $T$ (such that
the pulse area $\Omega T$ is an odd multiple of $\pi$), the
population of the initial state can be completely transferred into
the target state. Obviously, the main disadvantage of this method is
that the pulse area should be precisely designed.
Alternatively, in the second approach the population of the initial
state is transferred into the target state by utilizing various
population passage techniques~\cite{Bergmann}. Importantly, both the
adiabatical and non-adiabatical passages are insensitive to the areas of the applied pulses, and thus there is no need to design the exact
durations of the applied pulses. By this way, the desired quantum
manipulations for quantum computing could be realized more easily,
at least in principle.

Usually, the population passages are demonstrated by various
adiabatic manipulations, owing to its relative simplicity. For
example, one of the adiabatic passage technique called Stark-chirped
rapid adiabatic passage (SCRAP) has been attracting a lot of
attention in modern atomic, molecular and optical
physics~\cite{Rickes,Rangelov,Yatsenko}. Different from the other
adiabatical passages such as the stimulated Raman adiabatical
passage (STIRAP), the Stark shifts in the SCRAP caused by the
external pulses are beneficial to create the desired level-crossings
needed for population passages~\cite{Yatsenko}. Also, unlike the
technique in STIRAP with the assistants of various intermediate
states, the population transfers between the two selected quantum
states can be directly realized by just controlling the relative
intensity and delay time between the two applied pulses. Therefore, the
operations of SCRAPs are relatively simple and have been utilizing
in various quantum state engineerings based on population passages.
Certainly, the adiabatic requirement in the SCRAP is a fundamental
hurdle for its various potential applications. Under such a
limit, the applied pulse operations should be sufficiently slow such
that the evolution of the manipulated quantum state is adiabatic.
However, the practically-existing finite decoherence time of the
manipulated quantum system requires that the applied operations
should be sufficiently fast such that the operations could be
completed before the manipulated quantum superposed state
decoheres.
Therefore, relaxing the adiabatic condition is useful to
experimentally implement the fast population transfers for desired
quantum computing.

In this paper, by making use of the evolution-time insensitivity, we propose
the adiabatic and non-adiabatic population passages to implement the
quantum computation with electrons floating on the liquid helium.
In fact, since the pioneer work of Platzman and
Dykman~\cite{Platzman}, electrons floating on the liquid helium have
been served as one of the most hopeful candidates for implementing
quantum computation. In particular, due to the significantly long
decoherence time (e.g., estimated as $0.1 $ms for the electronic
qubit and $100$s for spin qubit~\cite{Lyon}) and the scalability, quantum
manipulations of surface-state electrons on liquid helium have attracted much attention in recent years~\cite{Dykman, Mostame, Collin}.
Note that, until now, almost all the proposals for quantum
manipulations of surface-state electrons on liquid helium are based
on the usual Rabi oscillation scheme, wherein the applied pulses
should be exactly designed to implement the desired quantum gate
operations. Given that the precise designs of all the applied pluses
(e.g., control the always-on inter-electron Coulomb interactions for
implementing the tunable two-qubit operation) are not practically
easy, here we propose an alternative approach to
implement the fundamental quantum logic operations by population
passages, wherein the durations of the applied pulses are not
required to be {\it precisely} designed.
In our proposal two pulses, one to chirp the qubit
state and one to drive the transition between the qubit states, are
enough to deterministically transfer the populations from one logic
state to the other. As a consequence, quantum logic gates with the
surface-state electrons on liquid helium can be implemented in a
evolution-time insensitive way by a pair of passage pulses. We
believe that the proposal could be realized experimentally, once the
single-quantum manipulations of surface-state electrons are
experimental demonstrated.

This paper is organized as follows. In Sec. II, we briefly review
the previous scheme of quantum computation with electrons floating
on liquid helium by using the evolution-time sensitive pulses (such
as Rabi oscillations). Our approach to implement the quantum logic
gates with these surface-state electrons by adiabatic and
nonadiabatic passages are introduced in Secs. III and IV,
respectively. Finally, we end with some discussions in Sec.
V.

\section{Quantum states of trapped electrons and their coherent controls
by exactly designed pulses}
\subsection{Hydrogen-like levels for electrons on liquid helium}
A typical experimental geometry of one electron being trapped above
the liquid helium is shown schematically in Fig.~1. Here, the
electron is sufficiently isolated from the outside world, and just
influenced by the irregularities of the liquid helium surface.
\begin{figure}[htbp]\centering
\includegraphics[scale=0.5]{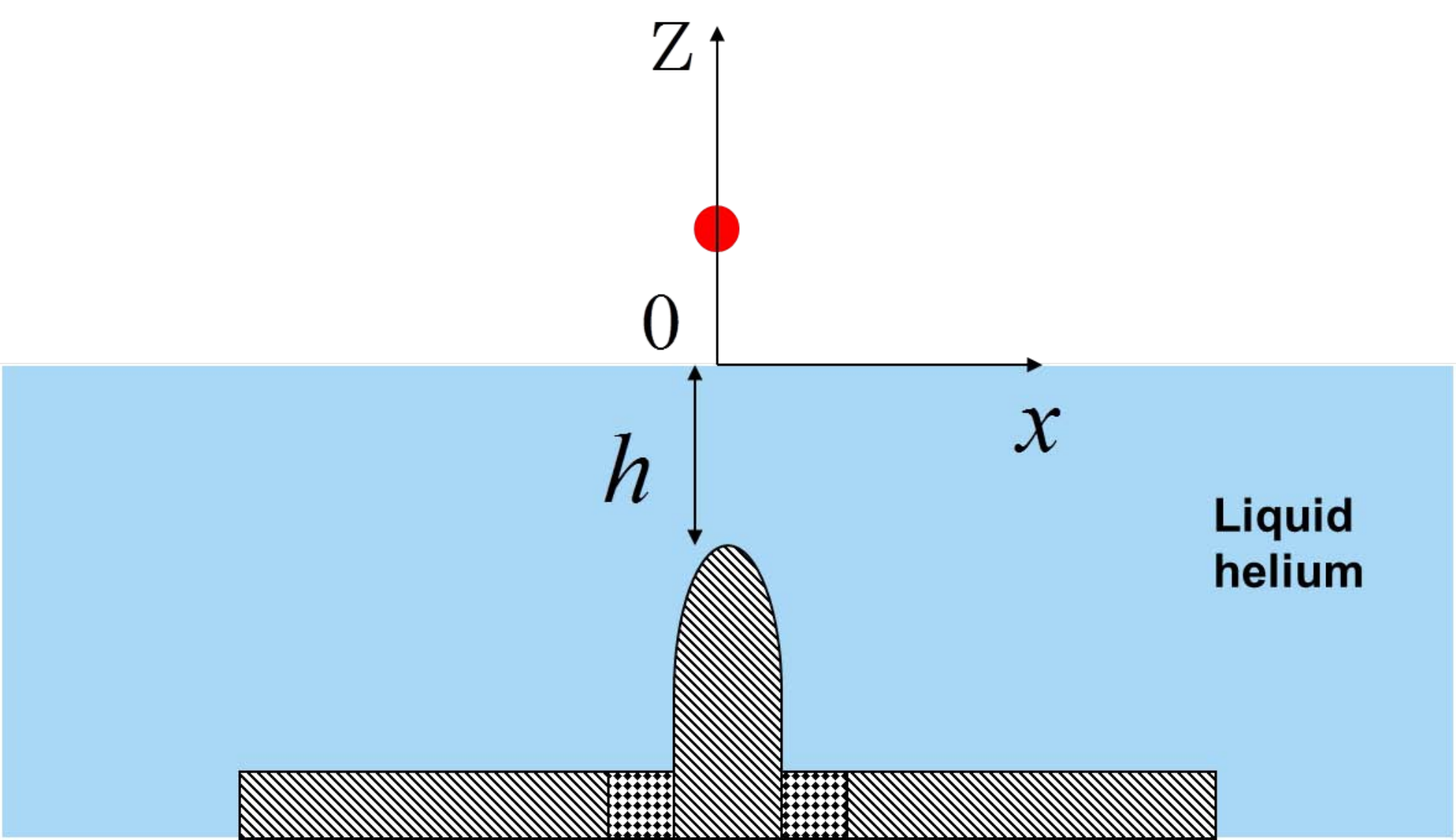}
\caption{(Color online) A sketch of an electron floating on the
surface of liquid helium and controlled by a
micro-electrode~\cite{Dykman}.}
\end{figure}
As the electron emerges in the vacuum over the liquid helium
surface, it will be attracted towards the surface by the
image-attractive force.
With a few angstroms of the surface, the electron encounters a
strong repulsive barrier along the $z$-direction (arising from the
Pauli exclusion principle), which prevents the electron from
penetrating into the liquid helium.
The potential well, formed by the sum of the image potential and the
repulsive barrier, supports a series of bound electronic states
which are very similar to those in the hydrogen atom.
Mathematically, such a 1D hydrogen-atom-like potential takes the
form~\cite{Zhang}
\begin{align}
V(z)=\left\{
        \begin{aligned}
          -\frac{\Lambda e^2}{z},~~~~~~&z>0\\
           V_0,                 ~~~~~~&z\leq 0
        \end{aligned}\right. \label{eq:1}
\end{align}
where $z$ is the coordinate of the electron perpendicular to the
interface and $e$ the charge of the electron. The coefficient
$\Lambda$ is $\Lambda\equiv (\epsilon-1)/[4(\epsilon+1)] \cong
0.0069$ with a helium dielectric constant of $\epsilon \cong 1.057$
~\cite{Gordon}. The height of the repulsive barrier $V_0$ has been
measured to be approximately $1~\rm eV$~\cite{Woolf}, which is a
sufficiently-strong obstacle such that the electron cannot drop into the liquid helium. Therefore, the Hamiltonian of a
single electron floating on the surface of liquid helium can be
simply written as
\begin{align}
\hat{H}=\frac{p^2}{2m}-\frac{\Lambda e^2}{z}.\label{eq:2}
\end{align}
By the usual finite difference method, we numerically solve the the
relevant spectrum problem related to the Hamiltonian \eqref{eq:2}
and find that the ground state and the lower two excited state
energies are $E_0=-0.65~\rm meV$, $E_1=-0.16~\rm meV$,
$E_2=-0.072~\rm meV$ respectively. The qubit is encoded by the two
lower states with the energy-splitting $\omega_{10}=(E_1-E_0)/\hbar\sim 2\pi\times 117~\rm GHz$. The
lower four states and their corresponding wavefunctions are shown
in Fig.~2.
Certainly, the energy-splittings of these bound states can be
controlled by the Stark field applied normally to the
surface~\cite{Grimes}.
\begin{figure}[htbp]
\includegraphics[scale=0.7]{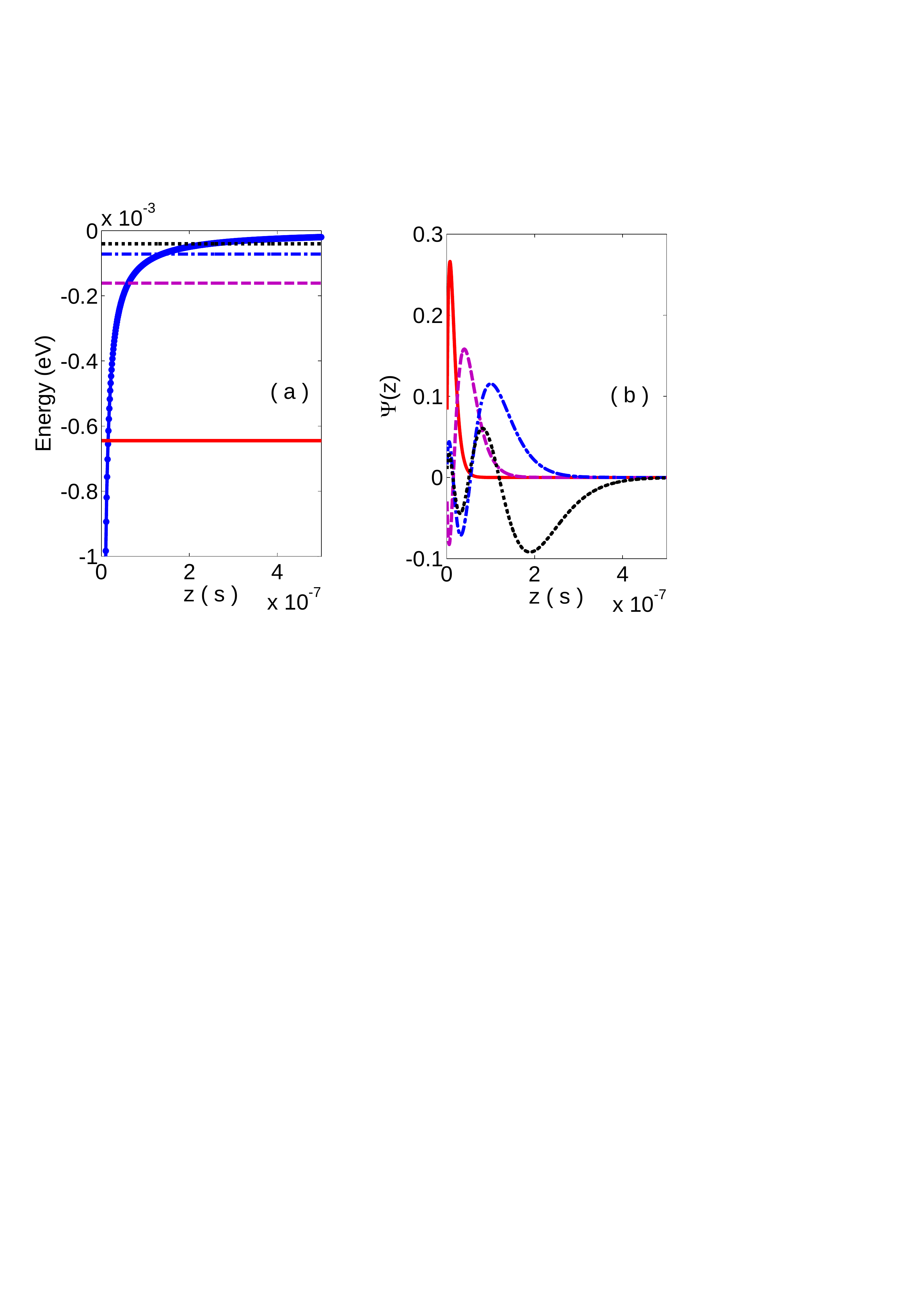}
\caption{(Color online) The lower four bound states and their
corresponding wave functions of a surface-state electron on liquid helium: (a) Lower four bound-state energy levels in the potential (solid blue line), and (b) The corresponding
eigenfunctions. The red, dashed purple, dotted-dashed blue and
dotted black lines denote the ground, the first excited, the second
excited and the third excited states, respectively.}
\end{figure}
\subsection{Rabi oscillations}
The most popular approach to implement the population transfer
between the selected qubit's states is by using the Rabi
oscillations, i.e., the population of one qubit's state undergoes a
sinusoidal variation. Here, the quantum transition between the
qubit's states is induced by applying a resonant driving
field~\cite{Bergmann}.
The Hamiltonian of the driven qubit reads
\begin{align}
\hat{H}_R(t)=\frac{\hbar\omega_0}{2}\hat{\sigma}_z+R(t)\hbar\hat{\sigma}_x,
\label{eq:3}
\end{align}
with $\omega_0=\omega_{10}$ being the transition frequency of the qubit and
$R(t)$ the controllable coupling between the qubit states;
$\hat{\sigma}_z$ and $\hat{\sigma}_x$ are the Pauli
operators, with
$\hat{\sigma}_x=|0\rangle\langle1|+|1\rangle\langle0|$ and
$\hat{\sigma}_z=|1\rangle\langle1|-|0\rangle\langle0|$. For the
resonant driving field $E_{ac}=\varepsilon(t)\cos(\nu t)$, the
induced coupling between the qubit's states takes the form
$R(t)=\Omega(t)\cos(\nu t)$, with $\nu=\omega_0$ and
$\Omega(t)=e\varepsilon(t)z_{10}/\hbar$ being the Rabi frequency.
Also, $\varepsilon(t)$ and $z_{10}=\langle 1| z |0\rangle$ are the
amplitude of the applied pulse and the matrix element of the
electric dipole moment respectively.

In the interaction picture defined by the unitary transformation
$\hat{U}_0=\exp[-i\omega_0\hat{\sigma}_z t/2]$, Eq. \eqref{eq:3} can
be transformed to the form
\begin{align}
\hat{H}_{\rm int}=\hat{U}_0^\dag R(t)\hat{\sigma}_x
\hat{U}_0=\frac{\Omega(t)}{2}\hbar\hat{\sigma}_x. \label{eq:4}
\end{align}
Formally, the time-evolution operator corresponding to this
Hamiltonian reads
\begin{align}
U(t)&=\exp\left(-\frac{i}{\hbar}\int_0^t\hat{H}_{\rm int}(t')dt'\right)\nonumber\\
    &=\cos(\frac{D(t)}{2})\rm
    I-i\sin(\frac{D(t)}{2})\hat{\sigma}_x,\label{eq:5}
\end{align}
with $D(t)=\int_0^t\Omega(t')dt'$ is the area of the pulse and $\rm I$ an unit matrix.
Obviously, if the qubit is initially prepared in the $|0\rangle$
state, then after a time $t$ the qubit will evolve into the
$|1\rangle$ state with a probability $P(t)=[1-\cos D(t)]/2$.
Therefore, in order to realize the complete inversion of the
population, the applied pulse must be precisely designed as a
$\pi$-pulse. Also, a precise ``$\pi/2$-pulse'' is required to
prepare the superposition state $(|0\rangle-i|1\rangle)/\sqrt{2}$.

\subsection{Precisely controlling durations for two-qubit gate}
A significant progress in quantum computing
is to achieve the single-qubit gate and the two-qubit gate. For the
present system the implementation of the desired two-qubit gate is not easy, as the interbit coupling is
always-on~\cite{Dykmanf}. Indeed, due to the always-on Coulomb
interaction,  the Hamiltonian of two nearest-neighbor electrons
moving along the $z$ direction can be well approximated as
\begin{align}
V_c(z_1,z_2)\approx\frac{e^2}{2d^3\cdot
       4\pi\epsilon_0}(z_1^2+z_2^2-2z_1z_2).
\end{align}
The Coulomb interaction between the electrons not only affects the
state-dependent shift of the energy of the neighbor electron but
also allows the resonant energy transfer from one electron to the
other~\cite{Platzman}.
In the computational basis, the Hamiltonian of the system can be
rewritten as
\begin{eqnarray}
\hat{H}_h=\sum_{j=1,2}\frac{\hbar\omega_j^0}{2}\hat{\sigma}_j^z
+V_c(z_1,z_2),
\end{eqnarray}
with $\hat{\sigma}_j^z= \rvert 1_j\rangle \langle 1_j\lvert-\rvert
0_j\rangle \langle 0_j\lvert$, and
\begin{eqnarray}
\hat{V}_c(z_1,z_2)&=&\zeta_1^z\hat{\sigma}_1^z+\zeta_2^z\hat{\sigma}_2^z+\zeta_1^x
                          \hat{\sigma}_1^x+\zeta_2^x\hat{\sigma}_2^x
                      +\zeta_{12}^{zz}\hat{\sigma}_1^z\hat{\sigma}_2^z
             +\zeta_{12}^{xx}\hat{\sigma}_1^x\hat{\sigma}_2^x
             +\zeta_{12}^{zx}\hat{\sigma}_1^z\hat{\sigma}_2^x+\zeta_{12}^{xz}
             \hat{\sigma}_1^x\hat{\sigma}_2^z.
\end{eqnarray}
In the above, $\omega_j^0$ is the transition frequency of the lower two
levels referring to the $j$th electron,
$\hat{\sigma}_j^x=\hat{\sigma}_j^+ +\hat{\sigma}_j^-$ with
$\hat{\sigma}_j^+=\rvert 1_j \rangle \langle 0_j\lvert$,
$\hat{\sigma}_j^-=\rvert 0_j \rangle \langle 1_j\lvert$; and
\begin{equation*}
\left\{\begin{aligned}
&\zeta_1^z=\frac{e^2}{16\pi\varepsilon_0d^3}(z_1^{00}+z_1^{11}-z_2^{00}-z_2^{11})(z_1^{11}-z_1^{00}),\\
&\zeta_2^z=\frac{e^2}{16\pi\varepsilon_0d^3}(z_2^{00}+z_2^{11}-z_1^{00}-z_1^{11})(z_2^{11}-z_2^{00}),\\
&\zeta_1^x=\frac{e^2}{8\pi\varepsilon_0d^3}(z_1^{00}+z_1^{11}-z_2^{00}-z_2^{11})z_1^{01},\\
&\zeta_2^x=\frac{e^2}{8\pi\varepsilon_0d^3}(z_2^{00}+z_2^{11}-z_1^{00}-z_1^{11})z_2^{01},
\end{aligned} \right.
\end{equation*}
and
\begin{equation*}
\left\{\begin{aligned}
&\zeta_{12}^{zz}=\frac{-e^2}{16\pi\varepsilon_0d^3}(z_1^{11}-z_1^{00})(z_2^{11}-z_2^{00}),\\
&\zeta_{12}^{xx}=\frac{-e^2}{4\pi\varepsilon_0d^3}z_1^{01}z_2^{01},\\
&\zeta_{12}^{zx}=\frac{-e^2}{8\pi\varepsilon_0d^3}(z_1^{11}-z_1^{00})z_2^{01},\\
&\zeta_{12}^{xz}=\frac{-e^2}{8\pi\varepsilon_0d^3}(z_2^{11}-z_2^{00})z_1^{01}.
\end{aligned} \right.
\end{equation*}
Here, $z_j^{11}$, $z_j^{00}$ and $z_j^{01}$ are the matrix elements
$\langle 1_j \lvert z_j \rvert 1_j \rangle$, $\langle 0_j \lvert z_j
\rvert 0_j \rangle$, and $\langle 1_j \lvert z_j \rvert 0_j
\rangle$, respectively. In the interaction picture defined by the
unitary
$\hat{U}_1(t)=\exp{[(-i/\hbar)t\sum_{j=1,2}\lambda_j\hat{\sigma}_j^z]}$
with $\lambda_j=\hbar\omega_j/2+\zeta_j^z$, the Hamiltonian of the
system becomes
\begin{align}
\hat{H}_{I}=&\zeta_{12}^{zz}\hat{\sigma}_1^z\hat{\sigma}_2^z+\sum_{j=1,2}\zeta_j^x
                        \left(e^{2it\lambda_j/\hbar}\hat{\sigma}_j^+
                  + e^{-2it\lambda_j/\hbar}\hat{\sigma}_j^-\right)\nonumber\\
+&\zeta_{12}^{xx}\left[e^{2it(\lambda_1+\lambda_2)/\hbar}\hat{\sigma}_1^+\hat{\sigma}_2^+
                  +e^{2it(\lambda_1-\lambda_2)/\hbar}\hat{\sigma}_1^+\hat{\sigma}_2^-\right.
           \left.+e^{-2it(\lambda_1-\lambda_2)/\hbar}\hat{\sigma}_1^-\hat{\sigma}_2^+
           +e^{-2it(\lambda_1+\lambda_2)/\hbar}\hat{\sigma}_1^-\hat{\sigma}_2^-\right]\nonumber\\
+&\zeta_{12}^{zx}\left(e^{2it\lambda_1/\hbar}\hat{\sigma}_1^z\hat{\sigma}_2^+ +
           e^{-2it\lambda_2/\hbar}\hat{\sigma}_1^z\hat{\sigma}_2^-\right)
           +\zeta_{12}^{xz}\left(e^{2it\lambda_1/\hbar}\hat{\sigma}_1^+\hat{\sigma}_2^z
           +e^{-2it\lambda_1/\hbar}\hat{\sigma}_1^-\hat{\sigma}_2^z\right).\label{eq:24}
\end{align}
Consequently, under the usual rotating-wave approximation, we have
\begin{align}
\overline{H}_{I}=\zeta_{12}^{zz}\hat{\sigma}_1^z\hat{\sigma}_2^z+\zeta_{12}^{xx}(\hat{\sigma}_1^+\hat{\sigma}_2^-
            +\hat{\sigma}_1^-\hat{\sigma}_2^+).\label{eq:25}
\end{align}
In the above derivation, we have assumed that
$\lambda_1=\lambda_2$ for simplicity. Obviously, the Hamiltonian
(10) yields the following two-qubit evolution
\begin{align}
\hat{U}&=e^{-i\hat{H}_It/\hbar}=\left(\begin{array}{cccc}
e^{-i\phi}         &0      &0           &0\\
0       &e^{i\phi}\cos{\xi}         &-i\sin{\xi}           &0\\
0        &-i\sin{\xi}       &e^{i\phi}\cos{\xi}          &0\\
0       &0         &0           &e^{-i\phi}
\end{array}\right),
\end{align}
$\text{with}\,\xi=t\zeta_{12}^{xx}/\hbar,\,\,\phi=t\zeta_{12}^{zz}/\hbar$.
This is a typical two-qubit i-SWAP gate. With such an universal
gate, assisted by arbitrary rotations of single qubits, any quantum
computing network could be constructed~\cite{Lloyds}.
If the two-qubit system is prepared beforehand in a pure state
$|01\rangle$ (or $|10\rangle$), then under the i-swap gate
operation, the state will evolve to
$\hat{U}|01\rangle=\exp{(i\phi)}\cos{\xi}|01\rangle-i\sin{\xi}|10\rangle$
(or
$\hat{U}|10\rangle=\exp{(i\phi)}\cos{\xi}|10\rangle-i\sin{\xi}|01\rangle$).
Again, the population inversion: $|01\rangle\longrightarrow
|10\rangle$ (or\,\, $|10\rangle\longrightarrow |01\rangle$) could be
exactly implemented by setting the evolution time exactly at
$t=\hbar\pi/(2\zeta_{12}^{xx})$.

In principle, quantum computation with surface-state electrons on
liquid helium can be implemented with the above universal gates.
However, for implementing these gates the evolution times of the
system should be exactly set. Any deviation of the desired
duration will decrease the fidelity of the expectable gate. In the
followings sections, we will show that the above quantum gates could
be implemented by using various evolution-time insensitive quantum
operations whose durations are not required to be exactly set.

\section{Quantum computation with trapped electrons on
helium by population passages I: adiabatic manipulations}
\subsection{The adiabatic population passage model: SCRAP}
Usually, the general logic gates in quantum computation are realized
by precisely designed resonant pulses. As it has been mentioned
above, in order to perform a one-qubit gate with an electron
floating on liquid helium, an exactly designed $\pi$-pulse should be
applied. However, due to the various fluctuations and operational
imperfections occuring in practice, it is not easy to precisely
design the applied pulses with exact durations.
Therefore, implementing quantum logic gates insensitive to evolution time is highly desirable in our context. In 2008, Wei et.~al.~\cite{Wei} proposed such an approach by using the Stark-chirped rapid adiabatic passage (SCPAP) technique. In this approach the dynamical Stark shift, induced by the applied pulse, is utilized to produce the required detuning chirp(s) of the qubit(s),
and the usual Rabi pulse is used to drive the quantum transition between the qubit's states. Once the orders of the applied pulses are properly set, the population of one qubit's states can be deterministically passaged to the other. As a consequence, the desired quantum gate is implemented.

The basic idea of the SCPAP is briefly reviewed as follows.
Generally, the time-dependent Hamiltonian of the driven two-level
system under the applied Stark-chirped and Rabi pulses, $\Omega(t)$
and $R(t)$, reads
\begin{align}
\hat{H}_1(t)=\frac{\hbar\omega_0}{2}\hat{\sigma}_z+\hbar
R(t)\hat{\sigma}_x+\hbar\Delta(t)
\frac{\hat{\sigma}_z}{2}.\label{eq:6}
\end{align}
In the interaction picture, this Hamiltonian reduces to
\begin{align}
\hat{H}_1'(t)&=\frac{\hbar}{2}\left(
          \begin{array}{cc}
          0          &\Omega(t)\\
          \Omega(t)  &2\Delta(t)
          \end{array}
              \right)
              =\frac{\hbar\varepsilon(t)}{2}\left(
          \begin{array}{cc}
          0          &\sin{2\theta(t)}\\
          \sin{2\theta(t)}  &2\cos{2\theta(t)}
          \end{array}
              \right),\label{eq:7}
\end{align}
with $\varepsilon(t)$$=$$\sqrt{\Delta^2(t)+\Omega^2(t)}$ and
$\theta(t)$$=$$\arctan[\Omega(t)/\Delta(t)]/2$. By solving the
instantaneous eigenvalue equation of the Hamiltonian (13), one obtains
the eigenvalues: $\lambda_\pm
=\hbar[\Delta(t)\pm\sqrt{\Delta^2(t)+\Omega^2(t)}]/2$, and the
relevant eigenvectors:
\begin{align}
&|\lambda_+(t)\rangle=
\sin\theta(t)|0\rangle+\cos\theta(t)|1\rangle,\\
&|\lambda_-(t)\rangle=\cos\theta(t)|0\rangle-\sin\theta(t)|1\rangle.
\end{align}
If the rate of the change of the Hamiltonian is slow enough, the so-called adiabatic condition,
\begin{align}
\eta=\frac{\left|\Omega(t)d\Delta(t)/dt-\Delta(t)d\Omega(t)/dt
\right|}{2[\Delta^2(t)+\Omega^2(t)]^{3/2}}\ll 1 \label{beta}
\end{align}
is satisfied, the system will stay at its instantaneous eigenstate.
And, there is no transition between the two instantaneous
eigenstates $|\lambda_-(t)\rangle$ and $|\lambda_+(t)\rangle$. This
implies that the system exists two independent adiabatic passages
defined by its two instantaneous eigenstates.

If the system is initially prepared at one of the adiabatic states,
it stays at this state during the adiabatic evolution but the populations of a diabatic basis $|0\rangle$ and $|1\rangle$ could be changed. Therefore, if the order of the applied pulses is right, the population of one state can
be completely passaged to that of another state. Certainly,
the relative phase of the computational basis depends on the evolution time. But, such a relative phase shift gate, which is
sufficiently produced by the Stark pulse, can be written as
$\hat{U}_p(\varrho)=\exp{(i\varrho|1\rangle\langle1|)}$ with
$\eta=-\int_{t_0}^{t_f}\Delta(t')dt'$. This phase shift gate implies
the evolution: $|0\rangle\rightarrow|0\rangle$ and
$|1\rangle\rightarrow e^{i\varrho}|1\rangle$. Therefore, the applied
stark pulse just changes the phase accumulation of the qubit state
$|1\rangle$, but does not destroy the population distribution
between the states $|0\rangle$ and  $|1\rangle$.

Obviously, the above adiabatic passage could be used to implement
the desired single-qubit quantum gates~\cite{Wei,Nie} for quantum
computing. Of course, if the population transfer between the qubit
states is not partial (but not completely inversion), then the
population can be controllably distributed at the qubit's states. As
a consequence, the relevant superposition of the states $|0\rangle$
and $|1\rangle$ can be obtained.

\subsection{Adiabatical population passages for single-qubit gates}
In the Sec.~II, we have shown that $z$-directional motion of an
electron, trapped on the surface of liquid helium, could be
effectively treated as hydrogenic-like atom. Therefore, the two
lower states, $|0\rangle$ and $|1\rangle$, could be utilized to
encode a qubit for storing quantum information.
In order to perform the single-qubit with such a surface-state
electron by using the SCRAP introduced previously, we apply a
resonant pump electric-field and a dc micro-pulse to the electrode
for coupling the qubit's states and inducing the desired Stark
shifts, respectively.
With these external drivings introduced, the Hamiltonian of a
floating electron on the liquid helium can be written as
\begin{align}
\hat{H}_1(t)=\hat{H}_0+\hat{H}'(t),\label{eq:17}
\end{align}
with
\begin{align}
\hat{H}_0=\frac{\hat{p}^2}{2m}-\frac{\Lambda
e^2}{z}=\sum_{i=0,1,2}E_i|i\rangle\langle i|,
\end{align}
and
\begin{align}
\hat{H}'(t)&=eE_{dc}(t)z+e\xi(t) z\cos{w_0 t}\nonumber\\
&=\left[eE_{dc}(t)+e\xi(t)\cos{w_0
t}\right]\sum_{i,j=0,1,2}z_{ij}|i\rangle\langle j|,
\end{align}
where $z_{ij}=\langle i| z | j\rangle$ and $z_{ij}=z_{ji}$. The
potential induced by the pump pulse is $V_{\rm pump}(t)=\xi(t)
\cos{\omega_0 t}$, where $\omega_0$ is
the splitting frequency of the two lower state and $\xi(t)$ denotes
intensity of the pump pulse. And $E_{dc}(t)$ leading to Stark shift of
the energy levels is induced by the applying a $dc$-electric field to
the electrode beneath the helium surface. Of course, the field
induced by the electrode will both affect parallel and vertical
motion of the electron to the surface. However, the parallel motion
is just a harmonic oscillator and the oscillation frequency ($20~\rm
GHz$) is sufficiently weak, and thus can be effectively neglected (
compared with the transition frequency of the qubit:
$\omega_0=2\pi\times 117~\rm GHz$).
Actually, the two pulses applied to implement the qubit operation
can both generate the Stark shifts and also couple the two states of
the qubit. However, in the interaction picture and under the usual
rotating-wave approximation, the Stark shift induced by the pump
pulse can be effectively ignored (compared with that induced by the
dc pulse), and also that coupling induced by the dc pulse is
negligible (compared with that induced by the applied pump pulse).

Under the unitary transformation $\hat{U}_0(t)=\exp{(-i\hat{H}_0
t/\hbar)}$, defined in the interaction picture, the Hamiltonian
\eqref{eq:17} can be rewritten as
\begin{align}
\Hat{H}_{i}(t)=\left(\begin{array}{ccc}
0 &\kappa z_{01} &0\\
\kappa z_{10} &eE_{dc}(t)(z_{11}-z_{00}) &0\\
0& 0,&eE_{dc}(t)(z_{22}-z_{00})
\end{array}\right),
\end{align}
where $\kappa=e\xi(t)/2$, $z_{00}=0.0115~\rm\mu m$,
$z_{11}=0.0461~\rm\mu m$, $z_{22}=0.1038~\rm\mu m$,
$z_{01}=z_{10}=-0.0043~\rm\mu m$, and $z_{12}=z_{21}=0.0142~\rm\mu
m$.
\begin{figure}[htbp]
\includegraphics[scale=0.7]{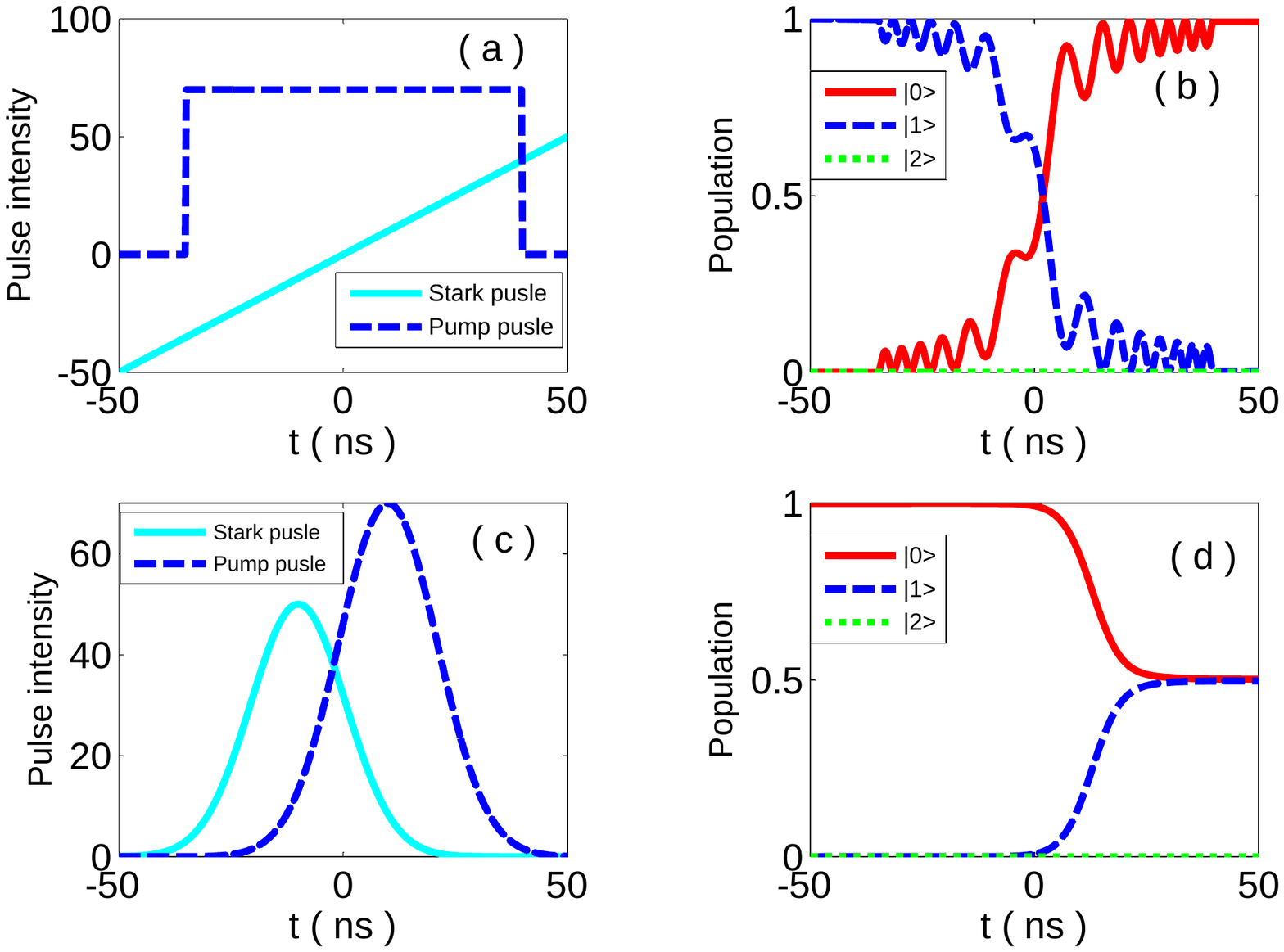}
\caption{(Color online) SCRAP-based population evolution of the
lower three states of the surface-state electron. The applied pulse
sequences are shown in (a), (c); and the corresponding population
transitions are shown in (b), (d), respectively. In (a) the two pulses
are designed with the linear forms $E_{dc}(t)=1\times 10^{9}t~\rm
V/m$ and $\xi(t)=70~\rm V/m~(-35~\rm n s\leq t\leq 40~\rm n
s,~\text{else where}~\xi(t)=0 ~\rm V/m)$. With this pulse sequence and
$\beta=0.13$, the electron initially prepared in the state
$|1\rangle$ completely passages to the state $|0\rangle$, see,
Fig.~3(b). It is clearly seen that the population leaking to the
upper state $|2\rangle$ is sufficient small that can be safely
neglected. In (c) two Gaussian,
$E_{dc}(t)=50\exp{[-(t+10)^2/(15)^2]}~\rm V/m$ and
$\xi(t)=70\exp{[-(t-10)^2/(15.5)^2]}~\rm V/m$ with $\beta=0.09$, are
designed to generate the superposition state
$(|0\rangle+|1\rangle)/\sqrt {2}$ from the initial state
$|0\rangle$. The relevant population changes are shown in Fig.~3(d).
It is also shown that during such a SCRAP, the population of the
state $|2\rangle$ is always sufficiently small and consequently any leakage of the qubit(s) can be neglected.}
\end{figure}

Fig.~3 shows the time evolutions of the populations in the lower
three states of surface-state electron during the designed SCRARs.
One can see that the usual $\hat{\sigma}_x$-rotation operation and
the Hadamard-like single-qubit gate could be implemented
adiabatically.
Specifically, when the pulses with the form dipicted in Fig.~3(a) are
applied to the electrode, the electron initially prepared in the
state $|1\rangle$ can be completely passaged to the state
$|0\rangle$ along the adiabatic path $|\lambda_-(t)\rangle$.
Conversely, if the electron initially prepared in the ground state
$|0\rangle$, then it is completely transferred to the first excited
state $|1\rangle$ along the adiabatic paths $|\lambda_+(t)\rangle$.
On the other hand, in order to utilize the SCRAP technique to
implement the Hadamard-like gate, we use two Gaussian pulses (rather
than the above linear pulses). The pulse sequence is indicated in
Fig.~3(c), wherein the Stark pulse is applied firstly and later the
pump pulse is applied, but the Stark pulse switches off prior to the
pump pulse. With such a pulse sequence, the initial evolution condition
$\theta(t_0)=\pi$ and the final one $\theta(t_f)=\pi/4$ are both
satisfied, thus the initial state $|0\rangle$ will evolve to the
superposition state $(|0\rangle+|1\rangle)/\sqrt {2}$ along the
adiabatic path $|\lambda_+(t)\rangle$, or to the state
$(|0\rangle-|1\rangle)/\sqrt{2}$ from the initial state
$|1\rangle$ via the population passage along the path
$|\lambda_-(t)\rangle$. Note again that, all these operations
require the orders of the applied pulses be properly designed, but not
the pulse durations.
Intuitively, the electron may populate to the higher state
$|2\rangle$ as the external pulse applied. However, as it is shown
in Fig. 3(b), the probability of the unwanted transition
between the states $|1\rangle$ and $|2\rangle$ is sufficiently small
and thus could be ignored.

It is emphasized that, the designed SCRAPs are really insensitive to
the durations of the pulses, as long as they are longer than, e.g.,
$\sim 100\,n$s. Also, our numerical solution confirms that the
desired adiabatic population transfers are still sufficiently
fast, e.g., within the time interval: $0.1~\rm \mu s$, which is
significantly shorter than the decoherence time $(\sim 0.1~\rm
ms$~\cite{Platzman}) of the system.

\subsection{Adiabatical population passages for two-qubit gates}
By precisely controlling the duration of the microwave pulse applied
via the patterned electrode (beneath the surface liquid helium), it
was shown~\cite{Platzman,Dykmanf,LEA} that a crucial two-qubit gate, i.e., the i-SWAP one, can be realized. In this subsection, we show
that this rigorous requirement of precisely designing the duration
of the applied pulse can be relaxed, and the same two-qubit gate
can be effectively realized via evolution-time insensitive
population passages.
\begin{figure}[htbp]
\centering
\includegraphics[scale=0.5]{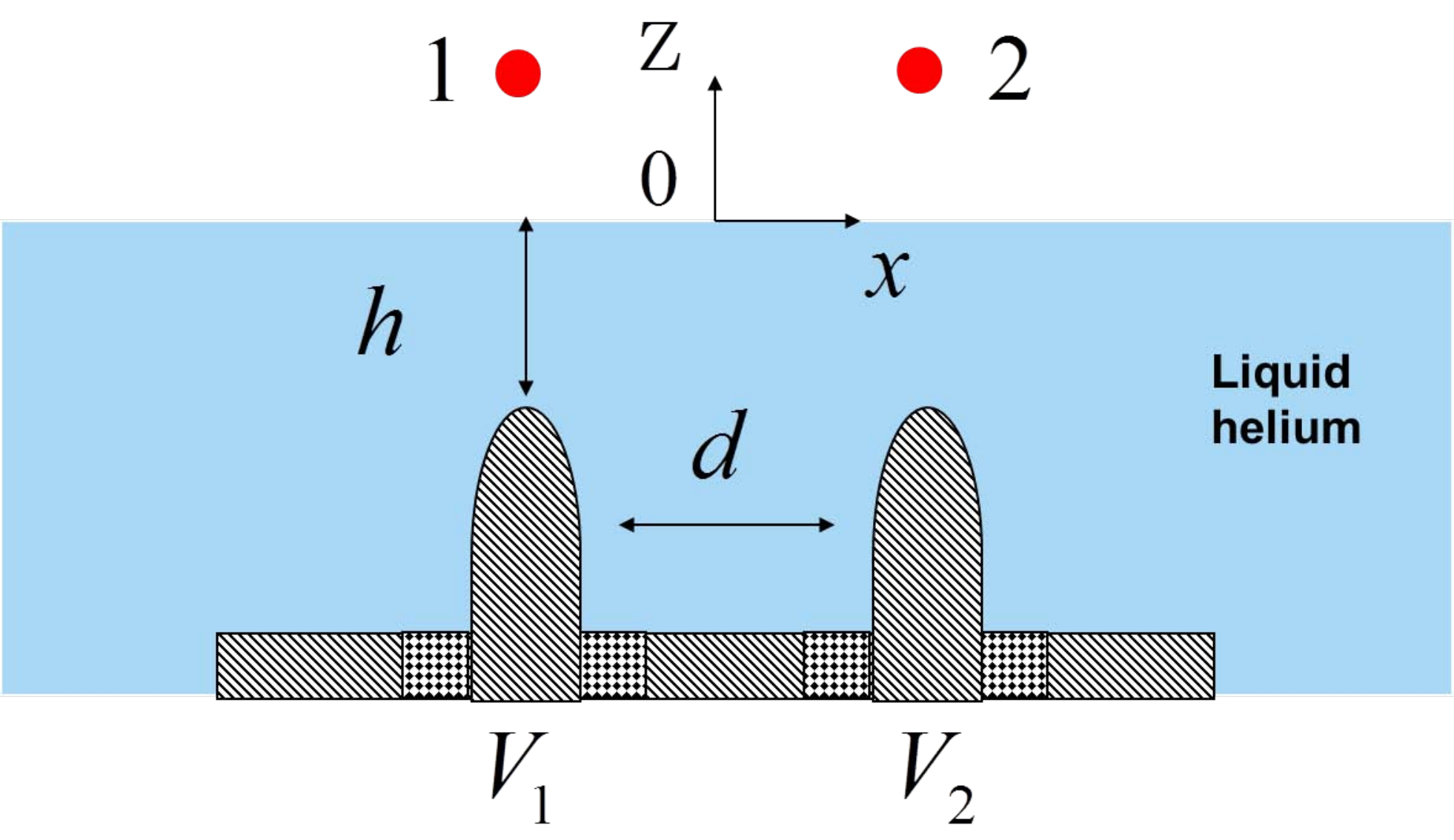}
\caption{(Color online) The geometry of a two-qubit system with two
electrons above the liquid helium and the micro-electrode. The
electrodes are designed to confine the electrons and induce the
Stark shifts to the electronic states. $V_1$ and $V_2$ are the
control potentials  of the two electrons, and $d\approx h=0.5~\rm\mu
m$.}
\end{figure}

We consider the two electrons (portrayed in Fig.~4) trapped in the
spatially-separated potentials $V_1$ and $V_2$, respectively.
Suppose that the electrons have been cooled to sufficiently-low
temperature and their surface-states are safely populated at the
ground state (if there is not any driving pulse). Also, the electric
fields applied to the electrons along the z-axis are different
(i.e., $\bar{E}_1\neq \bar{E}_2$) to make the electrons be
non-resonant. To realize the desired SCRAP, we now fix the
electric field applied to the first qubit and change that applied to
the second one.
The Hamiltonian describing the driven two-electron on the liquid helium can
be written as
\begin{align*}
\hat{H}_{\rm total}&=\sum_{i=1,2} \hat{H}_0^{(i)}(z_i)+\hat{H}_{\rm
       int}(z_1,z_2)+\hat{H}_{\rm stark}^{(2)}(t),
\end{align*}
with
\begin{align*}
\hat{H}_0^{(i)}&=\frac{p_i^2}{2m}-\frac{\Lambda
e^2}{z_i}+ez_i\bar{E}_i+\frac{e\bar{E}_jz_i}{\sqrt{8}},\,\,i\neq j=1,2,\\
\hat{H}_{\rm
       int}&=\frac{e^2}{4\pi\epsilon_0}
       \times\frac{z_1^2+z_2^2-2z_1z_2}{2d^3},
\end{align*}
and
\begin{align*} \hat{H}_{\rm
stark}^{(2)}(t)=\left(eE_{dc}^{(2)}(t)z_2
       +\frac{eE_{dc}^{(2)}(t)z_1}{\sqrt{8}}\right).
\end{align*}
Above, $\hat{H}_0^{(i)}$ is the non-perturbed Hamiltonian of each
electron, $\hat{H}_{\rm int}$ represents the Coulomb interaction
between the electrons (along the z-axis direction), and
$\hat{H}^{(2)}_{\rm stark}(t)$ is the chirp term of the second
electron implemented by applying a $dc$ electric field
$E_{dc}^{(2)}$ with changeable amplitude.
Since the energy-splitting between the ground state and the first
excited state is larger than the one between the first excited state
and the second excited state, the unwanted non-resonant transition
between the states $|1\rangle$ and $|2\rangle$ can be safely
neglected during the operations applied to the qubits. As a
consequence, in the computational basis the above total Hamiltonian
$\hat{H}_{\rm total}$ can be rewritten as
\begin{align}
\hat{H}_{\rm Int}(t)=\left(\begin{array}{cccc}
            \Delta_1         &0
               &0          &0  \\
            0         &\Delta_2
             &-2\alpha z_{1}^{01}z_{2}^{10}e^{-i\omega t}           &0  \\
            0         &-2\alpha z_{1}^{10}z_{2}^{01}e^{i\omega t}
                  &\Delta_3           &0  \\
            0         &0    &0            &\Delta_4
            \end{array}\right),\label{eq:21}
\end{align}
where $\alpha=e^2/(2d^3\times
            4\pi\epsilon_0),\,\omega=\omega_{10}^{(1)}-\omega_{10}^{(2)}$,\,\,
$\omega_{10}^{(i)}=(E_{1}^{(i)}-E_0^{(i)})/\hbar$, and
\begin{align*}
\Delta_1=&\alpha[(z_1^{00})^2+z_1^{01}z_1^{10}+(z_2^{00})^2+z_2^{01}
z_2^{10}-2z_1^{00}z_2^{00}]\nonumber
     +eE_{dc}^{(2)}z_2^{00}+\frac{eE_{dc}^{(2)}}{\sqrt{8}}z_1^{00},\\
\Delta_2=&\alpha[(z_1^{00})^2+z_1^{01}z_1^{10}+(z_2^{11})^2+z_2^{10}
z_2^{01}-2z_1^{00}z_2^{11}]\nonumber
     +eE_{dc}^{(2)}z_2^{11}+\frac{eE_{dc}^{(2)}}{\sqrt{8}}z_1^{00},
\end{align*}
and
\begin{align*}
\Delta_3=&\alpha[(z_1^{11})^2+z_1^{10}z_1^{01}+(z_2^{00})^2+z_2^{01}
z_2^{10}-2z_1^{11}z_2^{00}]
     +eE_{dc}^{(2)}z_2^{00}+\frac{eE_{dc}^{(2)}}{\sqrt{8}}z_1^{11},\\
\Delta_4=&\alpha[(z_1^{11})^2+z_1^{10}z_1^{01}+(z_2^{11})^2+z_2^{10}
z_2^{01}-2z_1^{11}z_2^{11}]
     +eE_{dc}^{(2)}z_2^{11}+\frac{eE_{dc}^{(2)}}{\sqrt{8}}z_1^{11},
\end{align*}
with $z_{ij}=\langle i|z|j\rangle$ and $z_{ij}=z_{ji}$.

It is obviously seen from the Hamiltonian \eqref{eq:21} that the
two-qubit state $|00\rangle$ (or $|11\rangle$) forms an invariant
subspace, i.e., if the two-qubit is initially prepared in the state
$|00\rangle$ (or $|11\rangle$), then it always populates in such a
state. On the otherhand, the states $|01\rangle$ and $|10\rangle$ together form an invariant subspace, i.e., only the transition between these states
is allowable, and in this invariant subspace the Hamiltonian of the
system reduces to
\begin{align}
\hat{H}_{\rm sub}(t)=\left(\begin{array}{cc}
         0  &-2\alpha z_{1}^{01}z_{2}^{10}e^{-i\omega t}\\
         -2\alpha z_{1}^{10}z_{2}^{01}e^{i\omega t}
                  &\Delta_3-\Delta_2
            \end{array}\right),\label{eq:9}
\end{align}
Fig.~5(a) exhibits how the population of one two-qubit state (e.g.,
$|01\rangle$) passages to another one (e.g., $|10\rangle$) along the
relevant adiabatic path. Here, the parameters used in the numerical
simulations are: $z_1^{00}=0.0115~\rm \mu m$, $z_1^{11}=0.0457~\rm
\mu m$, $z_1^{01}=z_1^{10}=-0.0043~\rm \mu m$, $z_2^{00}=0.0115~\rm
\mu m$, $z_2^{11}=0.0458~\rm \mu m$, $z_2^{01}=z_2^{10}=-0.0043~\rm
\mu m$, and $E_{dc}^{(2)}(t)=\gamma t$ with
$\gamma=1\times10^{9}\,V/(m\times s)$.
Fig.~5(b) clearly shows that the requirement of accurately
designing the interaction time between the qubits is removed. If the
electrons are initially prepared in the state $|01\rangle$, then the
population is adiabatically changed to the state $|10\rangle$ along
the red-line path in Fig.~5(a). Once the duration of the driving pulse is
set as $\tau_a>400$ns here, the population  is transferred
completely.

\begin{figure}[htbp]
\includegraphics[width=12cm,height=8cm]{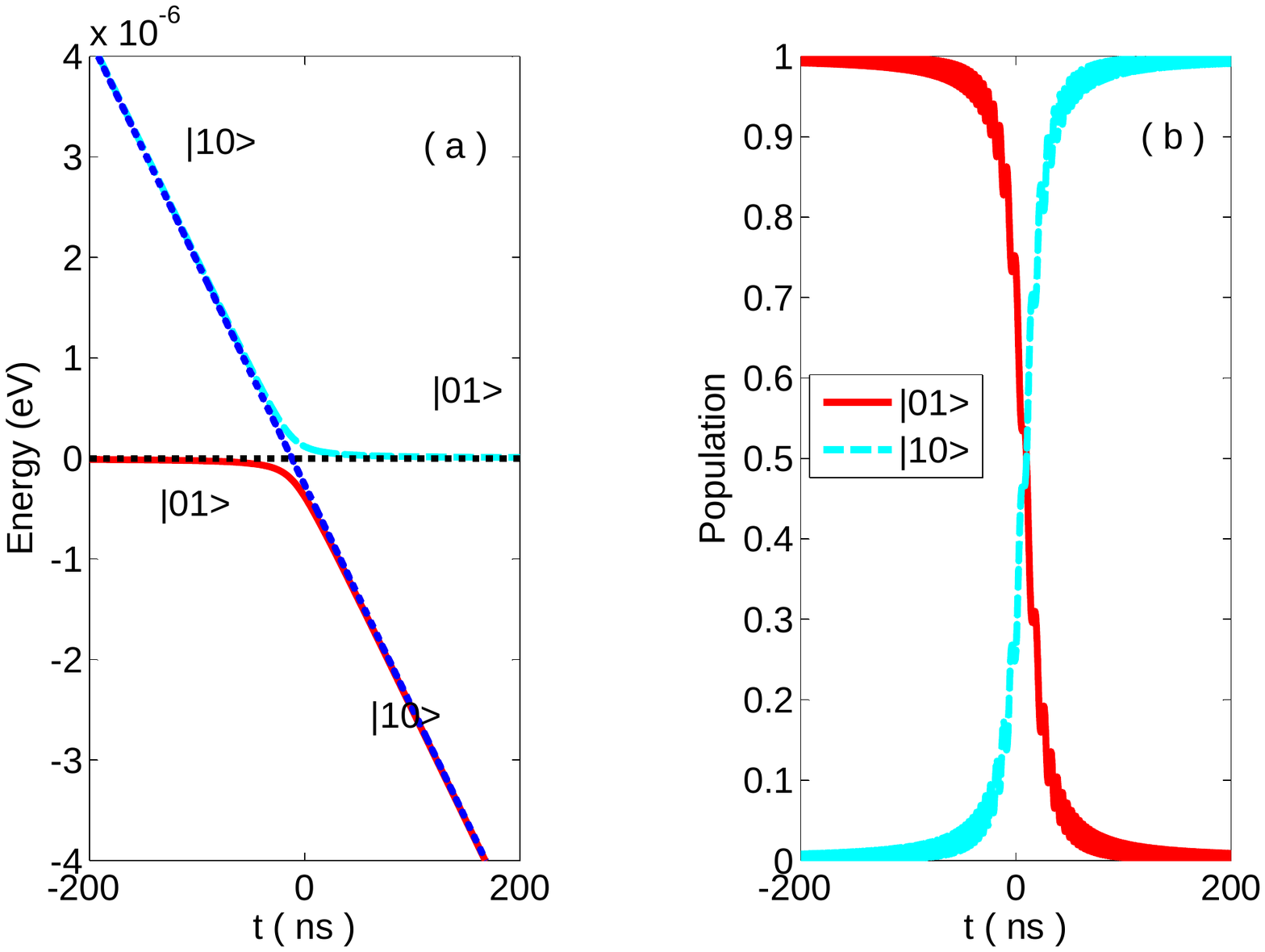}
\caption{(Color online) Population passages between the two
two-qubit states under the SCRAP. (a) Adiabatic passages, and (b)
the population transferred from the state $|01\rangle$ to the state
$|10\rangle$. The adiabatic parameter is $\beta=0.1$.}
\end{figure}

\section{Quantum computation with trapped electrons on helium by population passages II:
non-adiabatic manipulations}
\subsection{Evolution operator for a driven two-level system}
In the above SCRAP, the implementations of the usual quantum gates
are based on various adiabatic process along the independent paths.
Generally, these passages require that the evolutions of the driven
system should be significantly slow such that the adiabatic
conditions could be satisfied. However, the required adiabatic
condition is essentially limited by the defined decoherence time.
Therefore, relaxing the adiabatic condition required in the above
SCRAP technique is practically important for the realistic quantum
gates.

In order to implement the quantum computation beyond the adiabatic
limit, in this section we introduce a series of non-adiabatic passages of the
qubits' populations.
To this end, we begin with a generic time-driven two-level
system~\cite{Xian}
\begin{align}
\hat{H}(t)=[A(t)-\frac{1}{2}\hbar\dot{\omega}(t)]\hat{\sigma}_z
+B(t)\hat{\sigma}_x,\label{hh}
\end{align}
where $A(t)$, $B(t)$ and $\omega(t)$ are the controllable real
parameters. The dynamics of such a generic time-dependent system is described by the Schr$\rm \ddot{o}$dinger equation
\begin{align}
i\hbar\frac{\partial }{\partial
t}|\Psi(t)\rangle=\hat{H}(t)|\Psi(t)\rangle.\label{eq:10}
\end{align}
Suppose that the system is initially prepared in its ground state,
i.e., $|\Psi(t_0)\rangle=|0\rangle$, then the state after the
evolution time $t$ reads
$|\Psi(t)\rangle=\hat{U}(t)|\Psi(t_0)\rangle$ with $\hat{U}(t)$
being the evolution operator.
It obeys the follow equation
\begin{align}
i\hbar\frac{\partial }{\partial t}\hat{U}(t)=\hat{H}(t)\hat{U}(t),
\end{align}
with the formal solution~\cite{Ya},
\begin{align}
\hat{U}(t)=&\left(\begin{array}{cc}
\cos[\alpha(t)] &i\sin[\alpha(t)] \\
i\sin[\alpha(t)] &\cos[\alpha(t)]
\end{array}\right)\times
\left(\begin{array}{cc}
e^{i\gamma(t)} &0 \\
0 &e^{-i\gamma(t)}
\end{array}\right)
\times\left(\begin{array}{cc}
\cos[\beta(t)] &\sin[\beta(t)] \\
-\sin[\beta(t)] &\cos[\beta(t)]
\end{array}\right).
\end{align}
Here, the initial condition $\hat{U}(0)=1$ implies that,
$\alpha(0)=\beta(0)=\gamma(0)=0$, and the relevant time-dependent
parameters: $\alpha(t),\,\beta(t),$ and $\gamma(t)$, are determined
by the following differential equations:

\begin{equation}
\left\{\begin{aligned}
\dot{\alpha}&=-[A(t)-\frac{1}{2}\dot{\omega}(t)]\cos[2\alpha(t)]\tan
[2\beta(t)]-B(t),\\
\dot{\beta}&=[A(t)-\frac{1}{2}\dot{\omega}(t)]\sin[2\alpha(t)],\\
\dot{\gamma}&=-[A(t)-\frac{1}{2}\dot{\omega}(t)]\frac{\cos[2\alpha(t)]}
{\cos[2\beta(t)]}.
\end{aligned} \right.\label{eq:11}
\end{equation}

Once Eq. \eqref{eq:11} is solved, wave-function of the system at any
time can be obtained.

\begin{figure}[htbp]
\includegraphics[scale=0.7]{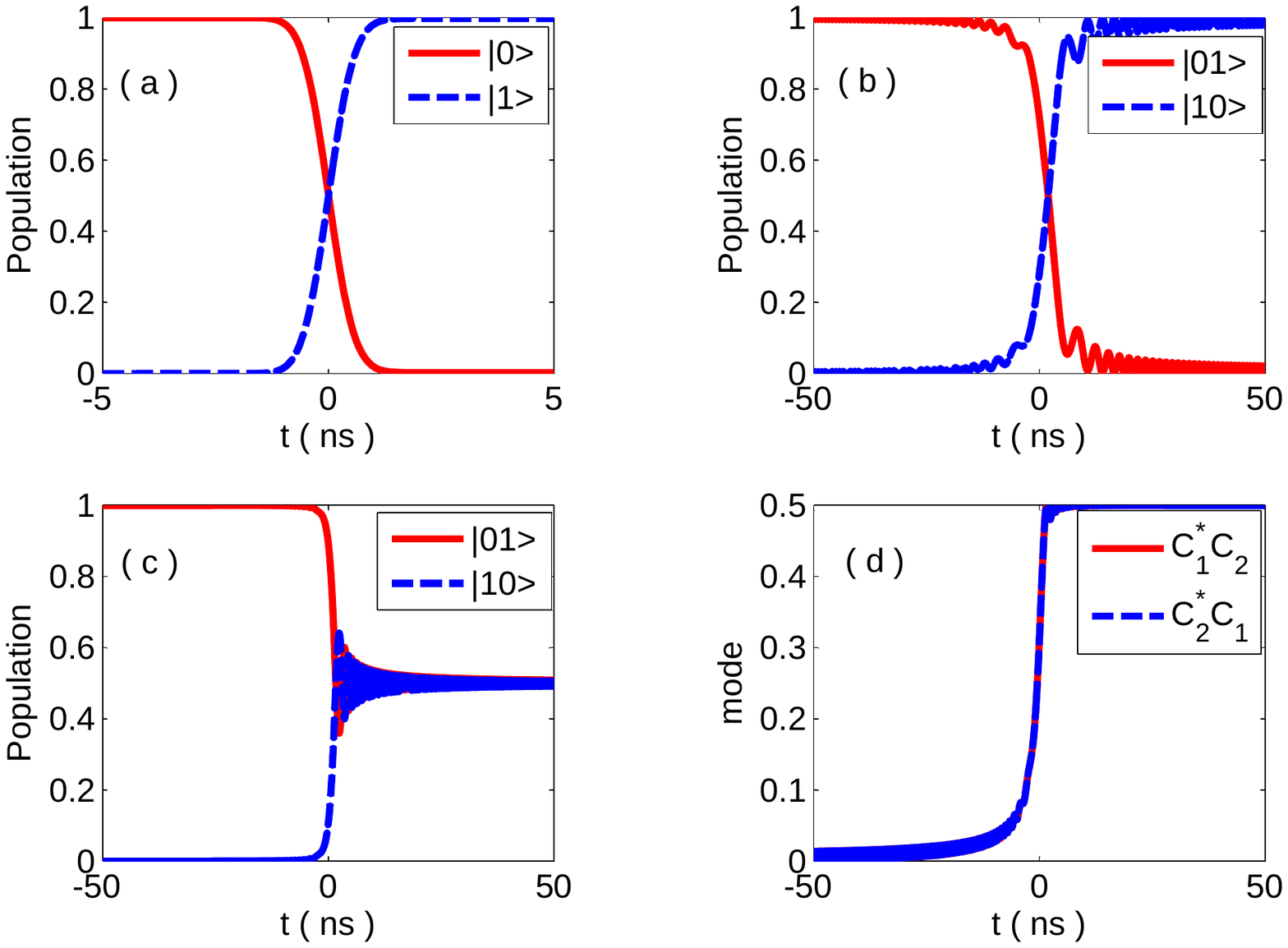}
\caption{(Color online) Non-adiabatic implementation of the usual
quantum gates and the preparation of Bell state. The
$\hat{\sigma}_x$-rotating gate shown in (a) is achievable by using
the Gaussian pulses $E_{dc}'(t)=10\exp{[-t^2/(1\times-10)^2]}~\rm
V/m$ and $E_{ac}'(t)=270\exp{[-t^2/(1\times-9)^2]}~\rm V/m$. The
$i$-swap gate is achieved by introducing a linear $dc$ micro-field
to the second electron: $E_{dc}'^{(2)}(t)=6\times10^9 t~\rm V/m$,
shown in (b). In (c) the equivalent-probability superposition of the
states $|01\rangle$ and $|10\rangle$ is given, after a fast population
passage process by applying the pulse $E_{dc}'^{(2)}(t)=28\times10^9
t~\rm V/m$. In (d) the non-zero real parts of the non-diagonal elements
of the density matrix of the generated superposed state are shown. This
confirms that the system is really prepared in a coherent
superposition state (rather than the mixed one).}
\end{figure}

\subsection{Non-adiabatical passage for the single-qubit gate}
The Hamiltonian describing the present fast-driven single qubit can
be expressed as
\begin{align}
\hat{H}_1(t)=\frac{eE_{dc}'(t)(z_{11}-z_{00})}{2}\hat{\sigma}_z
+\frac{e\xi'(t)z_{01}}{2}\hat{\sigma}_x,\label{eq:12}
\end{align}
where $E_{dc}'(t)$ is the dc micro-pulse (which induces the
desired stark shift) and $\xi'(t)$ is the amplitude of the
resonant pump electric-field (which is used to couple the qubit's
states).
By Runge-Kutta method we can numerically solve the relevant
Eq.~\eqref{eq:11} and then exactly determine the time-evolution
operator $\hat{U}(t)$. Consequently, if the qubit begins with the
ground state $|0\rangle$ (the excited state $|1\rangle$), then after
the time $t$, the probability of the population being transferred
completely to the excited state $|1\rangle$ (the ground state
$|0\rangle$) is
\begin{align}
P_1(t)&=1-|\langle 0|\hat{U}(t)|0\rangle|^2\nonumber\\
&=1-\left|\cos[\alpha(t)]\cos[\beta(t)]-i\sin[\alpha(t)]\sin[\beta(t)]\right|^2,
\end{align}
or
\begin{align}
P_0(t)&=1-|\langle 1|\hat{U}(t)|1\rangle|^2\nonumber\\
&=1-\left|i\sin[\alpha(t)]\cos[\beta(t)]-\cos[\alpha(t)]\sin[\beta(t)]\right|^2.
\end{align}
Fig.~6(a) shows that the $\hat{\sigma}_x$-rotation gate is
implemented by introducing two Gaussian pulses.
Note that the Gaussian pulses introduced here go against the
adiabatic condition \eqref{beta} (i.e., the adiabatic parameter is
$\beta=0.7$), and thus the population passages for implementing the
$\hat{\sigma}_x$-rotation gate are non-adiabatic. Indeed, in the
adiabatic basis: $|\lambda_-(t)\rangle$ and $|\lambda_+(t)\rangle$,
the Hamiltonian of the present driven qubit can be rewritten as
\begin{align}
\hat{H}_{nad}=\left(\begin{array}{cc}
 \lambda_-(t)   &-id\theta(t)/dt\\
 id\theta(t)/dt          &\lambda_+(t)
 \end{array}\right).
\end{align}
Obviously, the non-diagonal elements are not zero, and thus the
populations of the two adiabatic states ($|\lambda_-(t)\rangle$ and
$|\lambda_+(t)\rangle$) are oscillating before the desired
passages finish.

\subsection{Non-adiabatical passage for the two-qubit gate}

To implement the two-qubit i-swap gate, we need to achieve the
$\pi/2$-rotation between the states $|01\rangle$ and $|10\rangle$.
The dynamics for realizing such a driving is
\begin{align}
\hat{H}_{\rm sub}'(t)=\left(\begin{array}{cc}
         \Delta'_2 &-2\alpha z_{1}^{01}z_{2}^{10}e^{-i\omega t}\\
         -2\alpha z_{1}^{10}z_{2}^{01}e^{i\omega t}    &\Delta'_3
            \end{array}\right),
\end{align}
which is similar to that in Eq.~\eqref{hh} for the previously
adiabatic passage. Here,
\begin{align*}
\Delta_2'=&\alpha[(z_1^{00})^2+z_1^{01}z_1^{10}+(z_2^{11})^2+z_2^{10}
z_2^{01}-2z_1^{00}z_2^{11}]
     +eE_{dc}'^{(2)}z_2^{11}+\frac{eE_{dc}'^{(2)}}{\sqrt{8}}z_1^{00},\\
\Delta_3'=&\alpha[(z_1^{11})^2+z_1^{10}z_1^{01}+(z_2^{00})^2+z_2^{01}
z_2^{10}-2z_1^{11}z_2^{00}]
     +eE_{dc}'^{(2)}z_2^{00}+\frac{eE_{dc}'^{(2)}}{\sqrt{8}}z_1^{11}.
\end{align*}
The above Hamiltonian can be further simplified as
\begin{align}
\hat{H}_2=&\left(\frac{\Delta_2'-\Delta_3'}{2}-\frac{1}{2}\hbar\omega\right)
(|01\rangle\langle01|-|10\rangle\langle10|)-2\alpha(z_{1}^{01}z_{2}^{10}|01\rangle\langle10|+|10\rangle\langle01|).
\end{align}
Again, we can numerically solve the evolution operator corresponding
this Hamiltonian and then investigate the population transfers
between the states $|01\rangle$ and $|10\rangle$. Fig.~6(b) shows
clearly that, if the proper amplitude-adjustable $dc$ micro-field is
applied to the second electron, the desired $i$-swap gate can be
implemented non-adiabatically. The relevant adiabatic parameter is
$\beta=0.5$, and the time interval for realizing such a gate is
shortened to $\tau_{\rm na}> 100$ ns (which is significantly shorter
than $\tau_a>400$ns for the precious adiabatic passage.)

Furthermore, by designing the proper non-adiabatic passage the
two-qubit Bell state (the maximal entangled one of the two-qubit
system) can be fast generated deterministically. In fact, our
numerical results shown in Fig.~6(c) indicates that, an
equivalent-probability superposition of the states $|01\rangle$ and
$|10\rangle)$ is deterministically generated after the non-adiabatic
passages of the populations (with the adiabatic parameter
$\beta=2.15$). Also, we can prove that such a superposition is
coherent. Indeed, in the subspace spanned by the states $|01\rangle$
and $|01\rangle$, the evolution of the system can be generally
expressed as
\begin{align}
\psi_{\rm sub}(t)=C_1(t)|01\rangle +C_2(t)|10\rangle,
\end{align}
where $C_1(t)$ and $C_2(t)$ are the probabilistic amplitudes. Thus, the
reduced density matrix in this subspace reads
\begin{align}
\rho_{\rm sub}(t)=\left(\begin{array}{cc}
|C_1(t)|^2   &C_1^*C_2(t)\\
 C_2^*(t)C_1(t)         &|C_2(t)|^2
 \end{array}\right).
\end{align}
Suppose that the system is initially prepared in the state
$|01\rangle$, then the corresponding probability amplitudes (at any
time $t$) are
$C_1(t)=\{\cos[\alpha(t)]\cos[\beta(t)]-i\sin[\alpha(t)]\sin[\beta(t)]\}\exp{(i\gamma)}$
and $C_2(t)=\{i\sin[\alpha(t)]\cos[\beta(t)]
-\cos[\alpha(t)]\sin[\beta(t)]\}\exp{(i\gamma)}$.
Fig.~6(d) shows that the real parts of the parameters
$C_1^*(t)C_2(t)$ and $ C_2^*(t)C_1(t)$ are really not zero. This
indicates that the superposition of the states $|01\rangle$ and
$|10\rangle$ (deterministically generated by the above non-adiabatic
population passages) is coherent.

\section{Conclusions and Discussions}
In summary, we have investigated how to implement the single- and
two-qubit gates by using the population passage technique. We have shown
that the deterministic populations transfers between the selected
quantum states can be achieved by applying either adiabatic or
non-adiabatic pulses to the qubit(s). It is emphasized that,
differing from the approach to implement the quantum gates by using
precisely-designed pulses with exact durations, the quantum gates
implemented by the present population transfers are insensitive to
the durations of the applied pulses, whatever the relevant
evolutions are adiabatic or not. For the adiabatic passages the
driven qubit(s) evolves along the induced adiabatic paths. While,
under the non-adiabatic drivings the desired quantum gates could
still be deterministically implemented within the
significantly-short time. This is important to robustly overcome the
decoherence existed in the realistic systems. With the proposed
non-adiabatic passage technique, we have also shown that the Bell
state could be deterministically generated. 

Our generic proposal is demonstrated specifically with the
surface-state electrons on the liquid helium~\cite{Platzman}. This
system possesses several important advantageous features, e.g., it
could be conveniently manipulated by the electrodes beneath the
helium surface, the system could be easily fabricated and
integrated, and the qubit(s) can be simply addressed and read out,
etc.. Particularly, the surface-state electrons on the liquid helium
possess the sufficiently-long coherence times. For example, it can
reach to $0.1~\rm m s$ at $10~\rm mK$. While, the durations of the
proposed adiabatic passages to implemented the desired
single-qubit and two-qubit operations are just $100\rm ns$ and
$400\rm ns$, respectively. Furthermore, if the non-adiabatic
passages are applied, durations of $10\rm ns$ and $100\rm ns$ are
enough to realize the single-qubit and two-qubit gates,
respectively. Therefore, the feasibility of our proposal can be
tested relatively-easily with the surface-state electrons, although
it is still limited~\cite{Platzman} by the experimental challenge of
single-electron readouts.

\section*{Acknowledgments}

This work was supported in part by the National Science Foundation
grant Nos. 90921010, 11174373, the National Fundamental Research
Program of China through Grant No. 2010CB923104, and National Research Foundation and Ministry of Education,
Singapore (Grant No. WBS: R-710-000-008-271).

\end{document}